\documentclass[graybox,amsmath,amssymb]{svmult}

\usepackage{mathptmx}       
\usepackage{helvet}         
\usepackage{courier}        
\usepackage{type1cm}        
\usepackage{makeidx}         
\usepackage{graphicx}        
\usepackage{multicol}        
\usepackage[bottom]{footmisc}
\usepackage{bm}

\def\res{C_{\rm res}}
\newcommand{\n}{\mbox{\boldmath $\nabla$}}
\newcommand{\alt}{\mbox{\raisebox{-0.6ex}{$\stackrel{<}{\sim}$}}}

\makeindex             

\begin{document}

\title*{Quantum Measurements with Dynamically Bistable Systems}
\author{M. I. Dykman}
\institute{Mark Dykman \at Department of Physics and Astronomy,
Michigan State University, East Lansing, MI 48824, USA
\email{dykman@pa.msu.edu} }
%
%
\maketitle

\abstract{Periodically modulated nonlinear oscillators often display bistability of forced vibrations. This bistability can be used for new types of quantum measurements. They are based on switching between coexisting vibrational states. Since switching is accompanied by a large change of the amplitude and phase of forced vibrations, the measurements are highly sensitive. Quantum and classical noise plays dual role. It imposes a limitation on sensitivity in the familiar regime of a bifurcation amplifier. On the other hand, it makes it possible to use a bistable modulated oscillator in a new regime of a balanced dynamical bridge. We discuss the switching probabilities and show that they display scaling with control parameters. The critical exponents are found for different types of bifurcations and for different types of noise.}

\section{Introduction}
\label{sec:1}

\begin{figure}[h]
\sidecaption
\includegraphics[width=2.5in]{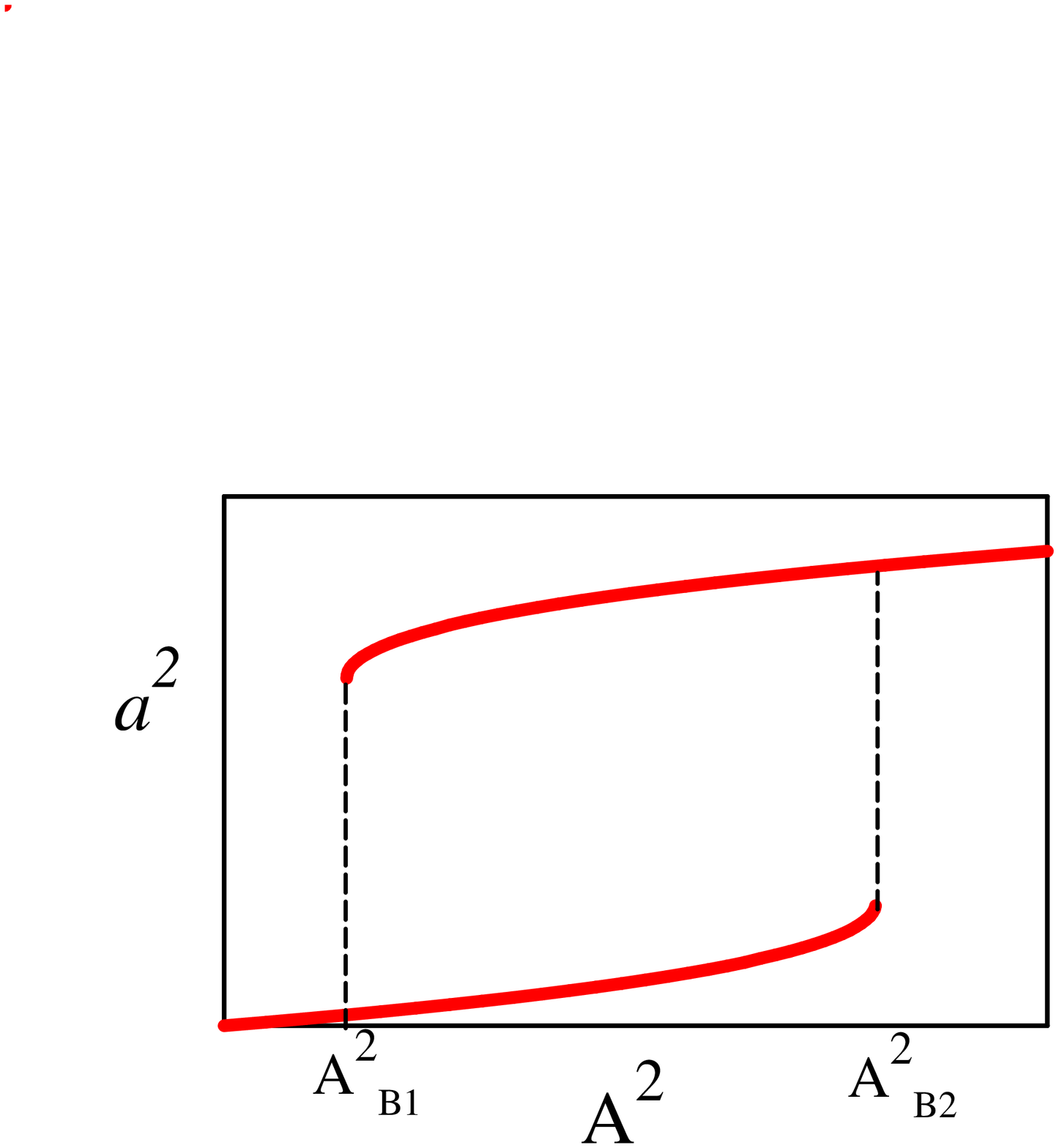}
\caption{Bistability of a nonlinear Duffing oscillator (\ref{eq:Duffing:general}). The solid lines show the squared amplitude of forced vibrations as a function of the squared amplitude of the driving force. At the bifurcational values $A^2_{\mathrm B1,2}$ one of the stable vibrational states disappears.}
\label{fig:1}       
\end{figure}

Bistability of vibrational states in modulated systems and fluctuation-induced switching between these states have attracted much attention recently. Experiments have been done on such diverse systems as electrons \cite{Lapidus1999} and atoms \cite{Gommers2005,Kim2006} in modulated traps, rf-driven Josephson junctions \cite{Siddiqi2004,Lupascu2006}, and nano- and micromechanical resonators  \cite{Aldridge2005,Badzey2005,Stambaugh2006,Almog2007}. These systems have small vibration damping, the quality factor may reach $10^4-10^5$. Therefore even a comparatively small resonant field can lead to coexistence of forced vibrations with different phases and amplitudes. This is illustrated in Fig.~\ref{fig:1}, which refers to a simple model relevant to many of the aforementioned experiments: an underdamped nonlinear classical oscillator driven close to resonance, with equation of motion
\begin{equation}
\label{eq:Duffing:general}
 \ddot q + \omega_0^2 q + \gamma q^3 + 2\Gamma \dot q=A\cos\omega_Ft.
\end{equation}
Here, $q$ is the oscillator coordinate, $\omega_0$ is its eigenfrequency, $\Gamma$ is the friction coefficient, $\Gamma\ll \omega_0$, and $\gamma$ is the nonlinearity parameter. The frequency of the modulating field $\omega_F$ is assumed to be close to $\omega_0$. In this case, for comparatively small modulation amplitude $A$, even where the oscillator becomes bistable its forced vibrations are nearly sinusoidal, $q(t)=a\cos(\omega_Ft+\phi)$ \cite{LL_Mechanics2004}.

The dynamic bistability of the oscillator is advantageous for measurements. The idea is to make the oscillator switch between the states depending on the value of the parameter to be measured. Switching leads to a strong change in the system that can be easily detected, leading to a high signal-to-noise ratio in a measurement. This has been successfully used for fast and sensitive measurements of the states of different types of  Josephson junction based qubits, including quantum non-demolition measurements \cite{Siddiqi2004,Lupascu2006,Siddiqi2006,Lupascu2007}.

So far the experiments were done in the bifurcation amplifier mode,
where the control parameter is swept through a bifurcation point (for
example, the field amplitude $A$ was swept through $A_{\mathrm B2}$,
see Fig.~\ref{fig:1}). The position of the bifurcation point, i.e.,
the value of $A$ where switching occurs, depends on the state of the
measured qubit. However, instead of happening for a certain $A$, as
expected from Fig.~\ref{fig:1}, switching happens at random within a
certain parameter range of $A$ near $A_{\mathrm B2}$.

The randomness of the switching field is a consequence of fluctuations in the oscillator. They lead to switching even before the control parameter reaches its bifurcational value. This is analogous to activated switching out of a potential well studied by Kramers \cite{Kramers1940}. However, in the case of an oscillator the stable states are not minima of a potential, and there is no static potential barrier that needs to be overcome. Switching of a modulated oscillator is an example of metastable decay of systems far from thermal equilibrium, the phenomenon of a broad interest.

Theoretical analysis of metastable decay requires developing methods for calculating the decay probability and finding out whether decay displays any universal system-independent features, like scaling dependence on the control parameters. For classical systems, scaling of the decay rate was indeed found for systems close to a bifurcation point, both in the cases of equilibrium \cite{Kurkijarvi1972,Victora1989,Garg1995a} and nonequilibrium systems \cite{Dykman1980,Dmitriev1986a,Tretiakov2005}. In the latter case a scaling crossover may occur as the system goes from the underdamped to overdamped regime while approaching the bifurcation point \cite{Dykman2005b}. Such crossover occurs also for quantum tunneling in equilibrium dissipative systems \cite{Caldeira1983}.

In this paper we study decay of metastable vibrational states in
quantum dissipative systems close to bifurcation points \cite{Dykman2007}. This is necessary for understanding the operation of a modulated oscillator in the regime of a bifurcation amplifier. We show that at low temperatures decay occurs via quantum activation. This is a specific process that has no analog in systems in thermal equilibrium \cite{Dykman1988a,Marthaler2006}. As tunneling, quantum activation is due to quantum fluctuations, but as thermal activation, it involves diffusion over an effective barrier separating the metastable state. As we show, near a bifurcation point quantum activation is more probable than tunneling even for $T\to 0$. We find that the decay rate $W$ scales with the distance to the bifurcation point $\eta$ as $|\ln W|\propto\eta^{\xi}$. The scaling exponent is $\xi=3/2$ for resonant driving, cf. Eq.~(\ref{eq:Duffing:general}). We also consider parametric resonance in a nonlinear oscillator and show that in this case $\xi=2$. In addition, $|\ln W|$ displays a characteristic
temperature dependence.

\section{Quantum Kinetic Equation for a Resonantly Driven Oscillator}
\label{sec:QKE}

The Hamiltonian of a resonantly driven nonlinear oscillator is
\begin{equation}
\label{eq:Hamiltonian_res} H_0(t)=\frac{1}{2}p^2 +
\frac{1}{2}\omega_0^2q^2 + \frac{1}{4}\gamma q^4 -qA\cos(\omega_Ft).
\end{equation}
The notations are the same as in equation of motion (\ref{eq:Duffing:general}), $p$ is the oscillator momentum. We assume that the detuning $\delta\omega = \omega_F -
\omega_0$ of the modulation frequency $\omega_F$ from the oscillator eigenfrequency $\omega_0$ is small and that $\gamma\;\delta\!\omega > 0$, which is necessary for bistability; for concreteness we set $\gamma>0$.

It is convenient to switch from $q,p$ to slowly varying operators
$Q,P$, using a transformation 
\begin{equation}
\label{eq:transformation}
q=\res(Q\cos\omega_F t+P\sin\omega_F
t), \quad p=-\res\omega_F (Q\sin\omega_F t - P\cos\omega_F t)
\end{equation}
with
$\res=(8\omega_F\delta\omega/3\gamma)^{1/2}$. The variables $Q,P$
are the scaled coordinate and momentum in the rotating frame. They are canonically conjugate,
\begin{equation}
\label{eq:commutator} [P,Q]=-i\lambda,\qquad \lambda
=3\hbar\gamma/8\omega_F^2\,\delta\!\omega.
\end{equation}
The parameter $\lambda$ plays the role of the effective Planck
constant. We are interested in the semiclassical case; $\lambda$ is
the small parameter of the theory, $\lambda \ll 1$.

\begin{figure}[h]
\includegraphics[width=3.8in]{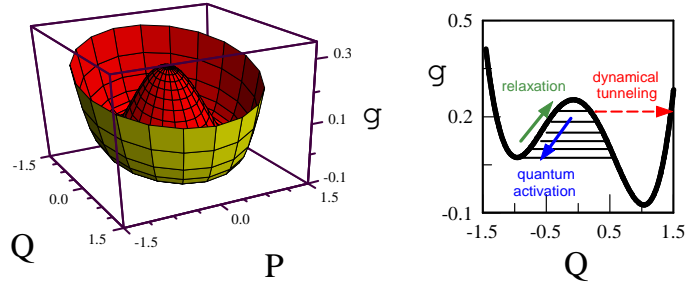}
\caption{Quasienergy of a resonantly driven nonlinear oscillator
$g(Q,P)$ (left panel) and its cross-section by the plane $P=0$ (right
panel). The plot refers to $\beta = 1/270$ in
Eq.~(\ref{eq:g_resonant}). Thin horizontal lines in the right
panel show (schematically) quasienergy levels for quantized motion around the
local maximum of $g(Q,P)$. In the presence of dissipation the states
at the local maximum and the minimum of $g(Q,P)$ become stable. They
correspond, respectively, to forced vibrations with small and large
amplitude $a$ in Fig.~\ref{fig:1}.  The arrows in the right panel show relaxation
to the state of small-amplitude vibrations, tunneling from this state
with constant quasienergy $g$, and quantum activation. The latter
corresponds to quantum diffusion over quasienergy away from the
metastable state, which accompanies relaxation
\cite{Dykman1988a,Marthaler2006}.}
\label{fig:2}       
\end{figure}

In the rotating wave approximation the Hamiltonian
(\ref{eq:Hamiltonian_res}) becomes $ H_0=
(\hbar/\lambda)\delta\!\omega\,\hat g$, with
\begin{eqnarray}
\label{eq:g_resonant}
\hat g\equiv g(Q,P)=\frac{1}{4}(Q^2+P^2-1)^2 - \beta^{1/2}Q, \qquad
\beta=3\gamma A^2/32\omega_F^3\left(\delta\omega\right)^3
\end{eqnarray}
(in the case $\gamma,\delta\omega <0$ one should use $g\to -g,
H_0\to -(\hbar/\lambda)\delta\!\omega\,g $). The function $g$ is shown in Fig.~\ref{fig:2}. It plays
the role of the oscillator Hamiltonian in dimensionless time $\tau =
t|\delta\omega|$. The eigenvalues of $\hat g$ give oscillator
quasienergies. The parameter $\beta$ in Eq.~(\ref{eq:g_resonant}) is the scaled
intensity of the driving field. For weak damping the oscillator is
bistable provided $0<\beta < 4/27$. The Heisenberg equation of motion for an arbitrary
operator $M$ is $\dot M\equiv dM/d\tau =-i\lambda^{-1}[M,g]$.

We will consider two major relaxation mechanisms of the oscillator:
damping due to coupling to a thermal bath and dephasing due to
oscillator frequency modulation by an external noise. Usually the
most important damping mechanism is transitions between neighboring
oscillator energy levels. They result from coupling linear in
the oscillator coordinate. Since the energy transfer is $\approx
\hbar\omega_0$, in the rotating frame the transitions look
instantaneous. Phenomenologically, the resulting relaxation may be described by a friction force proportional to velocity, as in (\ref{eq:Duffing:general}). Microscopically, such description applies in the case of Ohmic dissipation, i.e., coupling to Ohmic bath. However, we do not have to assume that dissipation is Ohmic. The only assumption needed for the further analysis is that the density of states of the reservoir weighted with the interaction be smooth in the frequency range, which is centered at $\omega_0$ and has a width that largely exceeds $\Gamma, |\delta\omega|$.

We will assume that the correlation time of the noise
that modulates the oscillator frequency is also short compared to
$1/|\delta\omega|$, so that the noise is effectively
$\delta$-correlated in slow time $\tau$. Then the quantum kinetic
equation is Markovian in the rotating frame. It has a familiar form (cf. \cite{DK_review84})
\begin{eqnarray}
\label{eq:QKE_general} \dot\rho\equiv
\partial_{\tau}\rho=i\lambda^{-1}[\rho,\hat g]-\hat{\Gamma}\rho\ -
\hat{\Gamma}^{\rm ph}\rho,
\end{eqnarray}
where $\hat\Gamma\rho$ describes damping,
\begin{eqnarray}
\label{eq:decay}
\hat{\Gamma}\rho=\Gamma|\delta\omega|^{-1}\left[(\bar{n}+1)(\hat
a^{\dagger}\hat a\rho-2\hat
a\rho\hat a^{\dagger}+ \rho\hat a^{\dagger}\hat a)
+\bar{n}(\hat a\hat a^{\dagger}\rho-2\hat
a^{\dagger}\rho\hat a +\rho\hat a\hat a^{\dagger})\right],
\end{eqnarray}
and $\hat{\Gamma}^{\rm ph}\rho$ describes dephasing,
\begin{equation}
\label{eq:dephasing} \hat{\Gamma}^{\rm ph}\rho=\Gamma^{\rm
ph}|\delta\omega|^{-1}\left[\hat a^{\dagger}\hat a,\left[\hat
a^{\dagger}\hat a,\rho\right]\right].
\end{equation}
Here, $\Gamma$ and $\Gamma^{\rm ph}$ are the damping and dephasing
rates, $\hat a= (2\lambda)^{-1/2}(Q+iP)$ is the oscillator lowering operator, and $\bar n = [\exp\left(\hbar\omega_0/k T\right)-1]^{-1}$ is the oscillator Planck number. In the classical mean-field limit one can obtain from (\ref{eq:QKE_general}), (\ref{eq:decay}) the same equation of motion as Eq.~(\ref{eq:Duffing:general}) in the rotating wave approximation.

In what follows we use dimensionless parameters
\begin{equation}
\label{eq:rates_ratio}\Omega=|\delta\omega|/\Gamma,\qquad
\kappa^{\rm ph} =\Gamma^{\rm ph}/\lambda\Gamma.
\end{equation}
We assume that $\kappa^{\rm ph}\alt 1$. This means that the intensity of phase fluctuations may be comparable to the intensity 
of quantum fluctuations associated with damping, which is $\propto
\lambda\Gamma$, see below, but that $\Gamma^{\rm ph}\ll \Gamma$.

Metastable decay was studied earlier for additively and
parametrically driven oscillators at $T=\kappa^{\rm ph}=0$ where
there is detailed balance
\cite{Drummond1980c,Drummond1989,Kryuchkyan1996}, and the lowest nonzero
eigenvalue of $\rho$ was studied numerically in Ref.~\cite{Vogel1988}.
However, the $T=\kappa^{\rm ph}=0$ solution is fragile. It can
change exponentially strongly already for extremely small
$T,\kappa^{\rm ph}$ \cite{Dykman1988a,Marthaler2006}. The analysis
\cite{Dykman1988a,Marthaler2006} revealed the mechanism of quantum
activation over a quasienergy barrier, but the results referred to
the case where the damping-induced broadening of quasienergy levels
is small compared to the typical interlevel distance. This condition
necessarily breaks sufficiently close to a bifurcation point where
the level spacing becomes small as a consequence of the motion
slowing down. 

\section{Wigner Representation}
\label{sec:Wigner}

The analysis of metastable decay near a bifurcation point can be conveniently done in the Wigner representation,
\begin{eqnarray}
\label{eq:Wigner_def} \rho_W( Q,P)=\int d\xi \,\E^{-i\xi
P/\lambda}\rho\left( Q+\frac{1}{2}\xi, Q-\frac{1}{2}\xi\right),
\end{eqnarray}
where $\rho(Q_1,Q_2)=\langle Q_1|\rho| Q_2\rangle$ is the density
matrix in the coordinate representation. Using
Eqs.~(\ref{eq:commutator})-(\ref{eq:Wigner_def}) one can formally
write the equation for $\rho_W$ as a sum of terms proportional to
different powers of the effective Planck constant $\lambda$,
\begin{eqnarray}
\label{eq:Liouville_compact} \dot\rho_W=-\n\cdot\left({\bf
K}\rho_W\right) +\lambda \hat L^{(1)}\rho_W +\lambda^2\hat
L^{(2)}\rho_W,
\end{eqnarray}
where ${\bf K} = (K_Q,K_P)$ and $\n = (\partial_Q,\partial_P)$.
Vector ${\bf K}$ determines evolution of the density matrix in
the absence of quantum and classical fluctuations,
\begin{eqnarray}
\label{eq:K_vector} K_Q=\partial_P g-\Omega^{-1}Q \qquad
K_P=-\partial_Q g-\Omega^{-1}P.
\end{eqnarray}
This evolution corresponds to classical motion
\begin{equation}
\label{eq:classical} \dot Q=K_Q,\qquad\dot P=K_P.
\end{equation}
The condition ${\bf K}={\bf 0}$ gives the values of $Q, P$ at the
stationary states of the oscillator in the rotating frame.

The term $\hat L^{(1)}$ in (\ref{eq:Liouville_compact})
describes classical and quantum fluctuations due to damping and
dephasing,
\begin{equation}
\label{eq:L1} \hat L^{(1)} = \Omega^{-1}\left[\left(\bar
n+\frac{1}{2}\right)\n^2 + \kappa^{\rm
ph}\left(Q\partial_P-P\partial_Q\right)^2\right].
\end{equation}
These fluctuations lead to diffusion in $(Q,P)$-space, as seen from
the structure of $\hat L^{(1)}$: this operator is quadratic in $\partial_Q, \partial_P$.

The term $\hat L^{(2)}$ in (\ref{eq:Liouville_compact})
describes quantum effects of motion of the isolated oscillator,
\begin{equation}
\label{eq:L2} \hat L^{(2)} =
-\frac{1}{4}\left(Q\partial_P-P\partial_Q\right)\n^2.
\end{equation}
In contrast to $\hat L^{(1)}$, the operator $\hat L^{(2)}$ contains
third derivatives. Generally the term $\lambda^2\hat L^{(2)}\rho_W$
is not small, because $\rho_W$ varies on distances $\sim\lambda$.
However, it becomes small close to bifurcation points, as shown
below.

\subsection{Vicinity of a Bifurcation Point}
\label{subsec:vicinity}

From (\ref{eq:K_vector}), (\ref{eq:classical}), for given
reduced damping $\Omega^{-1}$ the oscillator has two stable and one
unstable stationary state in the rotating frame (periodic states of
forced vibrations) in the range $\beta_B^{(1)}(\Omega) < \beta <
\beta_B^{(2)}(\Omega)$ and one stable state outside this range
\cite{LL_Mechanics2004}, with
\begin{equation} \label{eq:beta_B}
\beta_B^{(1,2)}=\frac{2}{27}\left[1+9\Omega^{-2}
\mp\left(1-3\Omega^{-2}\right)^{3/2}\right].
\end{equation}
At $\beta_B^{(1)}$ and $\beta_B^{(2)}$ the stable states with large
and small $Q^2+P^2$, respectively (large and small vibration
amplitudes), merge with the saddle state (saddle-node bifurcation).
The values of $Q,P$ at the bifurcation points 1, 2 are
\begin{equation}
\label{eq:equilibrium_bif}
Q_B=\beta_B^{-1/2}Y_B(Y_B-1), \qquad P_B=\beta_B^{-1/2}\Omega^{-1}Y_B \qquad (Y_B=Q_B^2+P_B^2),
\end{equation}
with
\begin{eqnarray}
\label{eq:positions_bif}
Y_B^{(1,2)}=\frac{1}{3}\left[2\pm(1-3\Omega^{-2})^{1/2}\right].
\end{eqnarray}

In the absence of fluctuations the dynamics of a classical system
near a saddle-node bifurcation point is controlled by one slow
variable \cite{Guckenheimer1987}. In our case it can be found by
expanding $K_{Q,P}$ in $\delta Q= Q-Q_B, \,\delta P=P-P_B$, and the
distance to the bifurcation point $\eta=\beta-\beta_B$. The function
$K_P$ does not contain linear terms in $\delta Q, \delta P$. Then,
from (\ref{eq:classical}), $P$ slowly varies in time for small
$\delta Q, \delta P, \eta$. 

On the other hand
\begin{eqnarray}
\label{eq:K_Q_B} K_Q\approx -2\Omega^{-1}\left(\delta Q- a_B\delta
P\right),\quad a_B=\Omega(2Y_B-1).
\end{eqnarray}
Therefore the relaxation time of $Q$ is $\Omega/2$, it does not
depend on the distance to the bifurcation point. As a consequence,
$Q$ follows $P$ adiabatically, i.e., over time $\sim\Omega$ it
adjusts to the instantaneous value of $P$.

\section{Metastable Decay near a Bifurcation Point}
\label{sec:decay_resonant}

The adiabatic approximation can be applied also to fluctuating
systems, and as we show it allows finding the rate of metastable decay. The approach is well known for classical systems described by the Fokker-Planck equation \cite{Haken2004}. It can be extended to the quantum problem by factoring $\rho_W$ into a normalized Gaussian distribution over $\delta\tilde{Q}=\delta Q-a_B\delta P$ and a function $\bar\rho_W(\delta P)$ that describes
the distribution over $\delta P$,
\[\rho_W\approx {\rm const}\times\exp\left[-2\frac{(\delta Q-a_B\delta P)^2}{\lambda\left(2\bar n+ 1\right)\left(1+a_B^2\right)}\right]\bar\rho_W(\delta P).\]
In the spirit of the adiabatic
approximation, $\bar\rho_W$ can be calculated disregarding small
fluctuations of $\delta\tilde{Q}$. Formally, one obtains an equation
for $\bar\rho_W$ by substituting the factorized distribution into
the full kinetic equation (\ref{eq:Liouville_compact}) and
integrating over $\delta\tilde{Q}$. This gives
\begin{eqnarray}
\label{eq:eq_bar_rho} \dot{\bar\rho}_W\approx
\partial_P\left[\bar\rho_W\partial_PU
+\lambda {\cal D}_B\partial_P\bar\rho_W \right],
\end{eqnarray}
where $U$ and ${\cal D}$ have the form
\begin{eqnarray}
\label{eq:U(x)_D} &&U=\frac{1}{3}b(\delta P)^3
-\frac{1}{2}\beta_B^{-1/2}\eta\delta P, \qquad \eta=\beta-\beta_B, \nonumber\\
&&{\cal D}_B=\Omega^{-1}\left[\left(\bar n+\frac{1}{2}\right)
+\frac{1}{2}\kappa^{\rm ph}(1-Y_B)\right]
\end{eqnarray}
with 
\[b=\frac{1}{2}\beta_B^{1/2}\Omega^2(3Y_B-2).\]
In (\ref{eq:eq_bar_rho}), (\ref{eq:U(x)_D}) we kept only the
lowest order terms in $\delta P, \beta-\beta_B,\lambda$. In
particular we dropped the term
$-\lambda^2Q_B\partial_P^3\bar\rho_W/4$ which comes from the
operator $\hat L^{(2)}$ in Eq.~(\ref{eq:Liouville_compact}). One can
show that, for typical $|\delta P|\sim |\eta|^{1/2}$, this term
leads to corrections $\sim \eta,\lambda$ to $\bar\rho_W$.

Eq.~(\ref{eq:eq_bar_rho}) has a standard form of the equation for
classical diffusion in a potential $U(\delta P)$, with diffusion
coefficient $\lambda{\cal D}_B$. However, in the present case the diffusion is due to quantum processes and the diffusion coefficient is $\propto \hbar$ for $T\to 0$. 

\subsection{Scaling of the Rate of Metastable Decay}
\label{subsec:scaling_resonant}

For $\eta b>0$ the potential $U$ (\ref{eq:U(x)_D})
has a minimum and a maximum. They correspond to the stable and
saddle states of the oscillator, respectively. The distribution $\rho_W$ has a diffusion-broadened peak at the stable state. Diffusion also leads to escape from the stable state, i.e., to metastable decay. The
decay rate $W$ is given by the Kramers theory \cite{Kramers1940},
\begin{eqnarray}
\label{eq:W_res} W=C\E^{-R_A/\lambda},\qquad
R_A=\frac{2^{1/2}|\eta|^{3/2}}{3{\cal D}_B|b|^{1/2}\beta_B^{3/4}},
\end{eqnarray}
with prefactor
$C=\pi^{-1}(b\eta/2)^{1/2}\beta_B^{-1/4}|\delta\omega|$ (in unscaled time $t$).

The rate (\ref{eq:W_res}) displays activation dependence on the
effective Planck constant $\lambda$.  The characteristic quantum
activation energy $R_A$ scales with the distance to the bifurcation
point $\eta=\beta-\beta_B$ as $\eta^{3/2}$. This scaling is
independent of temperature. However, the factor ${\cal D}_B$ in
$R_A$ displays a characteristic $T$ dependence. In the absence of
dephasing we have ${\cal D}_B=1/2\Omega$ for $\bar n\ll 1$, whereas
${\cal D}_B= kT/\hbar\omega_0\Omega$ for $\bar n\gg 1$. In the
latter case the expression for $W$ coincides with the result
\cite{Dykman1980}.

In the limit $\Omega\gg 1$ the activation energy (\ref{eq:W_res})
for the small-amplitude state has the same form as in the range of
$\beta$ still close but further away from the bifurcation point,
where the distance between quasienergy levels largely exceeds their
width \cite{Dykman1988a}. We note that the rate of tunneling decay
for this state is exponentially smaller. The tunneling is shown by the dashed line in the right panel of Fig.~\ref{fig:2}. The tunneling exponent for constant quasienergy scales as $\eta^{5/4}$ \cite{Dmitriev1986a,Serban2007}, which is parametrically larger than $\eta^{3/2}$ for small $\eta$ [for comparison, for a particle in a cubic potential (\ref{eq:U(x)_D}) the tunneling exponent in the strong-damping limit scales as $\eta$ \cite{Caldeira1983}].

For the large-amplitude state the quantum activation energy
(\ref{eq:W_res}) displays different scaling from that further
away from the bifurcation point, where $R_A\propto\beta^{1/2}$ for
$\Omega \gg 1$ \cite{Dykman1988a}. For this state we therefore
expect a scaling crossover to occur with varying $\beta$.

\section{Parametrically Modulated Oscillator}

The approach to decay of vibrational states can be extended to a
parametrically modulated oscillator that displays parametric resonance. The Hamiltonian of such an oscillator is
\begin{equation}
\label{eq:H_0_param(t)}
H_0(t)=\frac{1}{2}p^2+\frac{1}{2}q^2\left[\omega_0^2+F\cos(\omega_F
t)\right]+\frac{1}{4}\gamma q^4\, .
\end{equation}
When the modulation frequency $\omega_F$ is close to $2\omega_0$ the oscillator may have two stable
states of vibrations at frequency $\omega_F/2$ (period-two states)
shifted in phase by $\pi$ \cite{LL_Mechanics2004}. For $F\ll
\omega_0^2$ the oscillator dynamics is characterized by the
dimensionless frequency detuning $\mu$, effective Planck constant
$\lambda$, and relaxation time $\zeta$,
\begin{equation}
\label{eq:mu_and_lambda} \mu =
\frac{\omega_F(\omega_F-2\omega_0)}{F}, \quad \lambda =
\frac{3|\gamma|\hbar}{F\omega_F}, \quad
\zeta=\frac{F}{2\omega_F\Gamma}.
\end{equation}
As before, $\lambda$ will be the small parameter of the theory.

Parametric excitation requires that the modulation be sufficiently
strong, $\zeta > 1$. For such $\zeta$ the bifurcation values of
$\mu$ are
\begin{equation}
\label{eq:mu_B} \mu_B^{(1,2)}= \mp (1-\zeta^{-2})^{1/2}, \qquad
\zeta
> 1.
\end{equation}
If $\gamma>0$, as we assume, for $\mu<\mu_B^{(1)}$ the oscillator
has one stable state; the amplitude of vibrations at $\omega_F/2$ is zero. As $\mu$
increases and reaches $\mu_B^{(1)}$ this state becomes unstable and
there emerge two stable period two states, which are close in phase space for small $\mu-\mu_B^{(1)}$ (a supercritical pitchfork
bifurcation). These states remain stable for larger $\mu$. In addition, when
$\mu$ reaches $\mu_B^{(2)}$ the zero-amplitude state also becomes
stable (a subcritical pitchfork bifurcation). The case $\gamma < 0$
is described by replacing $\mu\to -\mu$.

The classical fluctuation-free dynamics for $\mu$ close to $\mu_B$
is controlled by one slow variable \cite{Guckenheimer1987}. The
analysis analogous to that for the resonant case shows that, in the
Wigner representation, fluctuations are described by one-dimensional
diffusion in a potential, which in the present case is quartic in
the slow variable. The probability $W$ of switching between the
period-two states for small $\mu - \mu_B^{(1)}$ and the decay
probability of the zero-amplitude state for small $\mu -
\mu_B^{(2)}$ have the form 
\begin{eqnarray}
\label{eq:W_par}W=C\exp(-R_A/\lambda), \qquad R_A=|\mu_B|\eta^2/2(2\bar n+1), \qquad \eta=\mu -
\mu_B
\end{eqnarray}
($\mu_B=\mu_B^{(1,2)}$). The corresponding prefactors are
$C_B^{(2)}=2C_B^{(1)}=2^{1/2}\pi^{-1}\Gamma\zeta^2|\mu_B||\mu -
\mu_B|$. Interestingly, dephasing does not affect the decay rate, to
the lowest order in $\mu-\mu_B$. This is a remarkable feature of quantum activation near bifurcation points at parametric resonance.

From (\ref{eq:W_par}), at parametric resonance the quantum
activation energy $R_A$ scales with the distance to the bifurcation
point as $\eta^2$. In the limit $\zeta \gg 1$ the same expression as (\ref{eq:W_par}) describes switching between period-two states
still close but further away from the bifurcation point, where the
distance between quasienergy levels largely exceeds their width. In
contrast, the exponent for tunneling decay in this case scales as
$\eta^{3/2}$ \cite{Marthaler2006}.

\section{Balanced Dynamical Bridge for Quantum Measurements}
\label{subsec:bridge}

Dynamic bistability of a resonantly driven oscillator can be used for quantum measurements in yet another way, which is based on a balanced bridge approach.
As a consequence of interstate transitions there is
ultimately formed a stationary distribution over coexisting stable
states of forced vibrations. The ratio of the state populations 
$w_{1},w_2$ is
\begin{equation}
\label{eq:population_ratio} w_1/w_2=W_{21}/W_{12}\propto
\exp[-(R_{A2}-R_{A1})/\lambda],
\end{equation}
where $W_{mn}$ is the switching rate $m\to n$.
For most parameter values $|R_{A1}-R_{A2}|\gg \lambda$, and then only one state is predominantly occupied. However, for a certain relation between $\beta$ and $\Omega$, where $R_{A1}\approx R_{A2}$, the
populations of the two states become equal to each other. This is a
kinetic phase transition \cite{Dykman1979a}. A number of unusual
effects related to this transition have been observed in recent
experiments for the case where fluctuations were dominated by
classical noise \cite{Almog2007,Stambaugh2006a,Chan2006}.

In the regime of the kinetic phase transition the oscillator acts
as a balanced dynamical bridge: the populations are almost equal with
no perturbation, but any perturbation that imbalances the activation energies leads to a dramatic change of $w_{1,2}$, making one of them practically equal to zero and the other to 1. Such a change can be
easily detected using, for example, the same detection technique as
in the bifurcation amplifier regime. An interesting application of a bridge is measurement of the statistics of white noise in quantum devices. The Gaussian part of the noise does not move a classical oscillator away from the kinetic phase transition. In contrast, the non-Gaussian part, which is extremely important and extremely hard to measure, does, and therefore can be quantitatively characterized.

\section{Discussion of Results}
\label{sec:discussion}

It follows from the above results that, both for resonant and
parametric modulation, close to bifurcation points decay of
metastable vibrational states occurs via quantum activation. The
quantum activation energy $R_A$ scales with the distance to the bifurcation point $\eta$ as $R_A\propto \eta^{\xi}$, with $\xi=3/2$ for resonant driving and $\xi=2$ for parametric resonance. The activation energy $R_A$ is smaller than the tunneling exponent.
Near bifurcation points these quantities become parametrically
different and scale as different powers of $\eta$, with the scaling exponent for tunneling ($5/4$ and $3/2$ for resonant driving and parametric resonance, respectively) being always smaller than for quantum activation.

The exponent of the decay rate $R_A/\lambda$ displays a characteristic dependence
on temperature. In the absence of dephasing, for $kT\gg
\hbar\omega_0$ we have standard thermal activation, $R_A\propto
1/T$. The low-temperature limit is described by the same expression
with $kT$ replaced by $\hbar\omega_0/2$. 

Our results show that quantum activation is a characteristic quantum
feature of metastable decay of vibrational states. Activated decay may
not be eliminated by lowering temperature. It imposes a limit on the
sensitivity of bifurcation amplifiers based on modulated Josephson
oscillators used for quantum measurements
\cite{Siddiqi2006,Lupascu2007}. At the same time, an advantageous
feature of dynamically bistable detectors is that they can be
conveniently controlled by changing the amplitude and frequency of the
modulating signal. Therefore the value of the activation energy may be
increased by adjusting these parameters. Another advantageous feature
of these detectors, which has made them so attractive, is that they
operate at high frequency and have a small response time, which can
also be controlled. The results of this work apply to currently
studied Josephson junctions, where quantum regime is within
reach. They apply also to nano- and micromechanical resonators that
can be used for high-sensitivity measurements in the regime of a
balanced dynamical bridge.

\begin{acknowledgement}
I am grateful to M. Devoret for the discussion. This
research was supported in part by the NSF through grant No.
PHY-0555346 and by the ARO through grant No.
W911NF-06-1-0324.
\end{acknowledgement}


\begin{thebibliography}{10}
\providecommand{\url}[1]{{#1}}
\providecommand{\urlprefix}{URL }
\expandafter\ifx\csname urlstyle\endcsname\relax
  \providecommand{\doi}[1]{DOI \discretionary{}{}{}#1}\else
  \providecommand{\doi}{DOI \discretionary{}{}{}\begingroup
  \urlstyle{rm}\Url}\fi

\bibitem{Lapidus1999}
L.J. Lapidus, D.~Enzer, G.~Gabrielse, Phys. Rev. Lett. \textbf{83}(5), 899
  (1999)

\bibitem{Gommers2005}
R.~Gommers, P.~Douglas, S.~Bergamini, M.~Goonasekera, P.H. Jones, F.~Renzoni,
  Phys. Rev. Lett. \textbf{94}(14), 143001 (2005)

\bibitem{Kim2006}
K.~Kim, M.S. Heo, K.H. Lee, K.~Jang, H.R. Noh, D.~Kim, W.~Jhe, Phys. Rev. Lett.
  \textbf{96}(15), 150601 (2006)

\bibitem{Siddiqi2004}
I.~Siddiqi, R.~Vijay, F.~Pierre, C.M. Wilson, M.~Metcalfe, C.~Rigetti,
  L.~Frunzio, M.H. Devoret, Phys. Rev. Lett. \textbf{93}(20), 207002 (2004)

\bibitem{Lupascu2006}
A.~Lupa\c{s}cu, E.F.C. Driessen, L.~Roschier, C.J.P.M. Harmans, J.E. Mooij,
  Phys. Rev. Lett. \textbf{96}(12), 127003 (2006)

\bibitem{Aldridge2005}
J.S. Aldridge, A.N. Cleland, Phys. Rev. Lett. \textbf{94}(15), 156403 (2005)

\bibitem{Badzey2005}
R.L. Badzey, G.~Zolfagharkhani, A.~Gaidarzhy, P.~Mohanty, Appl. Phys. Lett.
  \textbf{86}(2), 023106 (2005)

\bibitem{Stambaugh2006}
C.~Stambaugh, H.B. Chan, Phys. Rev. B \textbf{73}, 172302 (2006)

\bibitem{Almog2007}
R.~Almog, S.~Zaitsev, O.~Shtempluck, E.~Buks, Appl. Phys. Lett. \textbf{90}(1),
  013508 (2007)

\bibitem{LL_Mechanics2004}
L.D. Landau, E.M. Lifshitz, \emph{Mechanics}, 3rd edn. (Elsevier, Amsterdam,
  2004)

\bibitem{Siddiqi2006}
I.~Siddiqi, R.~Vijay, M.~Metcalfe, E.~Boaknin, L.~Frunzio, R.J. Schoelkopf,
  M.H. Devoret, Phys. Rev. B \textbf{73}(5), 054510 (2006)

\bibitem{Lupascu2007}
A.~Lupa\c{s}cu, S.~Saito, T.~Picot, P.C. De~Groot, C.J.P.M. Harmans, J.E.
  Mooij, Nature Physics \textbf{3}(2), 119 (2007)

\bibitem{Kramers1940}
H.~Kramers, Physica (Utrecht) \textbf{7}, 284 (1940)

\bibitem{Kurkijarvi1972}
J.~Kurkij\"arvi, Phys. Rev. B \textbf{6}, 832 (1972)

\bibitem{Victora1989}
R.~Victora, Phys. Rev. Lett. \textbf{63}, 457 (1989)

\bibitem{Garg1995a}
A.~Garg, Phys. Rev. B \textbf{51}(21), 15592 (1995)

\bibitem{Dykman1980}
M.I. Dykman, M.A. Krivoglaz, Physica A \textbf{104}(3), 480 (1980)

\bibitem{Dmitriev1986a}
A.P. Dmitriev, M.I. Dyakonov, Zh. Eksp. Teor. Fiz. \textbf{90}(4), 1430 (1986)

\bibitem{Tretiakov2005}
O.A. Tretiakov, K.A. Matveev, Phys. Rev. B \textbf{71}(16), 165326 (2005)

\bibitem{Dykman2005b}
M.I. Dykman, I.B. Schwartz, M.~Shapiro, Phys. Rev. E \textbf{72}(2), 021102
  (2005)

\bibitem{Caldeira1983}
A.O. Caldeira, A.J. Leggett, Ann. Phys. (N.Y.) \textbf{149}(2), 374 (1983)

\bibitem{Dykman2007}
M.I. Dykman, Phys. Rev. E \textbf{75}(1), 011101 (2007)

\bibitem{Dykman1988a}
M.I. Dykman, V.N. Smelyansky, Zh. Eksp. Teor. Fiz. \textbf{94}(9), 61 (1988)

\bibitem{Marthaler2006}
M.~Marthaler, M.I. Dykman, Phys. Rev. A \textbf{73}(4), 042108 (2006)

\bibitem{DK_review84}
M.I. Dykman, M.A. Krivoglaz, \emph{Soviet Physics Reviews} (Harwood Academic,
  New York, 1984), vol.~5, pp. 265--441

\bibitem{Drummond1980c}
P.D. Drummond, D.F. Walls, J. Phys. A \textbf{13}(2), 725 (1980)

\bibitem{Drummond1989}
P.D. Drummond, P.~Kinsler, Phys. Rev. A \textbf{40}(8), 4813 (1989)

\bibitem{Kryuchkyan1996}
G.Y. Kryuchkyan, K.V. Kheruntsyan, Opt. Commun. \textbf{127}(4-6), 230 (1996)

\bibitem{Vogel1988}
K.~Vogel, H.~Risken, Phys. Rev. A \textbf{38}(5), 2409 (1988)

\bibitem{Guckenheimer1987}
J.~Guckenheimer, P.~Holmes, \emph{Nonlinear Oscillators, Dynamical Systems and
  Bifurcations of Vector Fields} (Springer-Verlag, New York, 1987)

\bibitem{Haken2004}
H.~Haken, \emph{Synergetics: Introduction and Advanced Topics}
  (Springer-Verlag, Berlin, 2004)

\bibitem{Serban2007} I. Serban, F. Wilhelm, Phys. Rev. Lett. \textbf{99}(13), 137001 (2007)

\bibitem{Dykman1979a}
M.I. Dykman, M.A. Krivoglaz, Zh. Eksp. Teor. Fiz. \textbf{77}(1), 60 (1979)

\bibitem{Stambaugh2006a}
C.~Stambaugh, H.B. Chan, Phys. Rev. Lett. \textbf{97}(11), 110602 (2006)

\bibitem{Chan2006}
H.B. Chan, C.~Stambaugh, Phys. Rev. B \textbf{73}, 224301 (2006)

\end{thebibliography}

\end{document}